# Optimizing nucleation layers for the integration of ferroelectric HZO on CVD-grown graphene

Suzanne Lancaster
*Namlab gGmbH*
Dresden, Germany
suzanne.lancaster@namlab.com

Adrián Gudín
*IMDEA Nanociencia*
Madrid, Spain
adrian.gudin@imdea.org

Iciar Arnay
*IMDEA Nanociencia*
Madrid, Spain
iciar.arnay@imdea.org

Thomas Mikolajick
*Namlab gGmbH*
*IHM, TU Dresden*
Dresden, Germany
thomas.mikolajick@namlab.com

Stefan Slesazeck
*Namlab gGmbH*
Dresden, Germany
stefan.slesazeck@namlab.com

Ruben Guerrero
*IMDEA Nanociencia*
Madrid, Spain
*University of Castilla-La Mancha*
Ciudad Real, Spain
ruben.guerrero@imdea.org

Paolo Perna
*IMDEA Nanociencia*
Madrid, Spain
paolo.perna@imdea.org

*Abstract*- Direct integration of ferroelectric $Hf_{0.5}Zr_{0.5}O_2$ (HZO) on the inert surface of graphene is challenging. Here, using nucleation layers to promote atomic layer deposition of HZO was investigated. Different metals were deposited as nucleation layers via dc sputtering. Ta, which oxidizes in air to form a sub-stoichiometric oxide, was compared to Pt, which offers a more stable electrode. For thicker interlayers, Ta leads to unstable switching behavior of the HZO film. Conversely, at smaller thicknesses, a higher Pr can be achieved with an oxidized Ta interlayer. In both cases, Pt offers higher endurance. The choice of interlayer may strongly depend on the required application.

## I. Introduction

The combination of ferroelectrics with 2D materials such as graphene has strong potential for application in spin-orbitronics [1]. Large perpendicular magnetic anisotropy (PMA) [2] and Rashba-type (i.e. electrically tunable) Dzyaloshinskii-Moriya interactions [3,4] have been demonstrated in graphene/ferromagnet/heavy metal (Gr/FM/HM) stacks. Gr is an optimal spin channel material displaying several micrometers long spin-relaxation length at room temperature [5] and enhanced spin to charge conversion and spin Hall effect when interfaced with heavier atoms [6-8] This further motivates the integration of high-k dielectrics on Gr channels [9] for gate-tunable spintronics. The integration of ferroelectrics such as $Hf_{0.5}Zr_{0.5}O_2$ (HZO) on Gr channels opens up the possibility for exploiting different device concepts for adoption as non-volatile FE memories [10] with this novel material stack, for robust, fast and power-efficient next-generation memory devices.

While both atomic layer deposition (ALD) or physical vapor deposition (PVD) could be employed to deposit ferroelectric HZO, PVD bears the risk of damage to the Gr by either physical sputtering or oxidation. Further, PVD films typically require high temperatures for the crystallization anneal [11] which is necessary for most HZO films in order to achieve ferroelectricity. To avoid intermixing in our Co/Pt bilayers, the thermal budget for the deposition and anneal is strictly limited [12]. Therefore, several methods were investigated for enabling the ALD of ferroelectric (FE) HZO films on chemical vapor deposition (CVD)-grown Gr, despite the inherent inertness of the Gr surface. Direct deposition was investigated, but found to occur primarily on defect sites, which precludes the possibility of depositing a conformal film [13]. Surface treatment methods could be further investigated to enable direct deposition [14,15], but they need to fit strict requirements in terms of both chemistry and temperature. Instead, using a nucleation layer was investigated to enable conformal deposition on the Gr surface.

## II. Sample Fabrication

Graphene was deposited on Pt(111) templates via ethylene CVD at 1025 K under ultra-high vacuum. Co was deposited as a FM and intercalated through the graphene via annealing at a carefully controlled temperature; full details can be found in [12]. After deposition of the seed layer using either ALD or physical deposition methods (discussed further below), HZO was deposited via ALD with

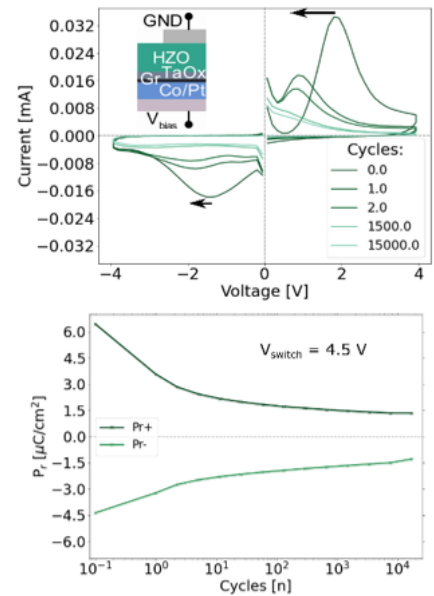

Figure 1: (a) Switching curves with cycling for 11 nm HZO on a 2nm TaOx layer on Gr/Co/Pt (cycling at 4.5 V, 100 kHz); (b) change in Pr as a function of cycles

*DOI: 10.1109/ISAF51494.2022.9870121*

alternating cycles of HfO₂ and ZrO₂ at 250°C. After deposition of either Al₂O₃ (ALD, TMA and H₂O at 250°C) or TiN (dc sputtering of Ti under 4 sccm N₂ flow), a spike anneal (5 s at 500 °C) was applied to promote crystallization into the ferroelectric orthorhombic Pca2₁ phase without any intermixing of the underlying metallic stack. By utilizing a thin nucleation layer, either in-situ ALD deposited or ex-situ using a physical deposition method, a ferroelectric film could be formed. Finally, capacitor structures were deposited by first removing the crystallization layer and depositing dot structures with 50-200 μm in diameter of Ti/Pt (10/25 nm).

## III. RESULTS

Various nucleation layers were investigated. While spin-orbit effects can be tuned through a metallic layer [16], applications for field-effect control of Gr require an interlayer with a thickness below its screening length. This requirement can be met by the adoption of either a dielectric or very thin metallic interlayer. Comparing different interlayers provides a roadmap for the integration of HZO films on Gr for different applications and should help to inform future investigations in this direction.

First, an in-situ ALD-deposited dielectric nucleation layer (Al₂O₃) was investigated, which was found to lead to a poor HZO film quality, due to non-uniform surface coverage and the formation of large defects [13]. Sputtered Ta was then investigated, where the metal was unintentionally oxidized via exposure to atmosphere over long periods (several days) during transfer between process steps. For a thick oxidized Ta layer (2 nm TaOx), the HZO layer shows both FE behaviour and a slower breakdown than was seen for HZO on Al₂O₃ seed layers. However, the resulting films show very rapid fatigue and a negative imprint, i.e. a shift of the coercive voltage $V_c$ to more negative voltages (fig. 1a), with the Pr declining from the first switching cycle (fig. 1b). This is related to a migration of charges to one interface, leading to domain pinning [17], and/or a field-induced phase change [18]. The quality of Gr was proven after the ALD deposition process by perpendicular magnetic anisotropy (PMA) measurements. Co/Pt demonstrates PMA up to Co thicknesses of 4 nm in the presence of graphene [2]. Magnetic hysteresis measurements were found to hardly change before or after the ALD process [13].

In order to confirm the impact of oxygen scavenging in substoichiometric, thicker TaOx interlayers, a comparison was made with an inert metal (Pt), where 3 nm of each metal was sputtered on FM/HM templates. Figure 2 shows the field cycling behaviour of HZO on TaOx vs. Pt layers. The applied voltage was always chosen to maximize Pr without early device breakdown (< 10⁴ cycles). In the case of thicker layers, using Pt rather than Ta improves both Pr and endurance. Furthermore, the immediate fatigue observed for TaOx interlayers was not present. We relate this to the inert Pt electrode blocking oxygen scavenging at the bottom electrode during cycling.

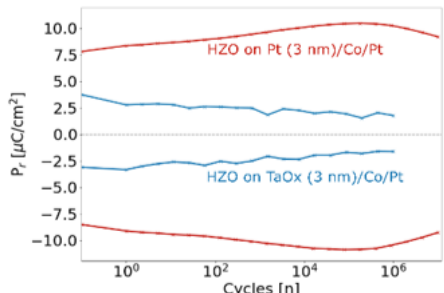

Figure 2: Pr with cycling of HZO on TaOx (3 nm)/Co/Pt (4.5 V cycling, 100 kHz) and Pt (3 nm)/Co/Pt (4.8 V cycling, 100 kHz)

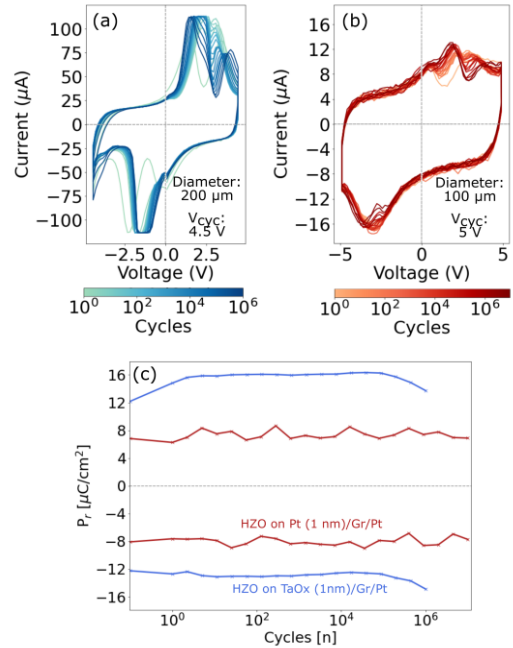

Figure 3: IV switching curves with cycling for (a) HZO on TaOx (1 nm)Gr /Pt and (b) HZO on Pt (1 nm)/Gr /Pt; (c) comparison of Pr with cycling for each film. Switching voltages are indicated in the IV plots.

Figure 3 shows a comparison of the switching curves (3a, 3b) and ferroelectric properties (3c) of HZO on thinner interlayers (1 nm) of TaOx and Pt on Gr/Pt. In this case, we see that reducing the thickness of Ta has a strong positive impact on both Pr and endurance. We attribute this to the full oxidation of the thin Ta layer in atmosphere, which improves the stability of the stack after the crystallization anneal and with electric field cycling. A fully oxidized Ta layer will remove less oxygen during cycling and generate fewer charges which can then migrate and lead to domain pinning in the case of the thicker Ta layer.

Conversely, the thickness of the Pt layer doesn't have a strong influence on either property, since oxygen transfer during the crystallization anneal and cycling should be relatively independent of the thickness with an inert electrode. These films don't show the slight improvement with cycling seen for thicker Pt layers (compare fig. 2 & 3c). Further, the switching (fig 3b) showed a larger coercive voltage $V_c$ than for the 3 nm Pt interlayer, and even at a slightly higher cycling voltage of 5 V, the switching peaks were cut off. At even higher voltages, the film endurance dropped considerably. On the other hand, films deposited on Pt consistently show a good endurance (> 10⁷ cycles) compared to those on Ta. The origin of the higher $V_c$ is still to be investigated. One possibility is that the Pt layer is not fully closed, which could also lead to damage of the underlying Gr. This should be confirmed via perpendicular magnetic anisotropy measurements on full stacks including Co. For thin TaOx interlayers, no such $V_c$ increase was observed until the interlayer thickness was further reduced to 0.5 nm.

## IV. CONCLUSIONS

To conclude, we have investigated the feasibility of different nucleation layers for depositing ferroelectric HZO on graphene. Both Pt and Ta thin films can be employed, depending on the intended application. Thin (1 nm) Ta interlayers lead to remanent polarization values, 2Pr, of up to 28 μC/cm$^2$ with an endurance of 10$^6$ cycles. On the other hand, Pt layers show a 2Pr of 16-20 μC/cm$^2$ which varies less with interlayer thickness, and an endurance $> 10^7$ cycles. From here, further parts of the process flow can be modified to improve the FE quality, for example, the oxygen content in the films, or the annealing conditions. Finally, the consistency of perpendicular magnetic anisotropy measurements before and after ALD confirms the minimal impact of FE deposition on the integrity of the underlying metallic stacks and, crucially, the Gr layer [13].

ACKNOWLEDGMENT

The authors acknowledge funding from the FLAG-ERA JTC 2019 grant SOgraphMEM through the partner's national research agencies AEI (Spain, PCI2019-111867-2) and DFG (Germany, MI 1247/18-1).